\documentclass[sigconf]{acmart}

\AtBeginDocument{%
  \providecommand\BibTeX{{%
    \normalfont B\kern-0.5em{\scshape i\kern-0.25em b}\kern-0.8em\TeX}}}

\setcopyright{acmcopyright}
\copyrightyear{2018}
\acmYear{2018}
\acmDOI{XXXXXXX.XXXXXXX}

\acmConference[MSR 2022]{MSR '22: Proceedings of the 19th International Conference on Mining Software Repositories}{May 23–24, 2022}{Pittsburgh, PA, USA}

\acmPrice{15.00}
\acmISBN{978-1-4503-XXXX-X/18/06}

\usepackage{algorithmic}
\usepackage{graphicx}
\usepackage{textcomp}
\usepackage{xcolor, colortbl}
\usepackage{booktabs}
\usepackage{multirow}
\usepackage{xspace}
\usepackage{caption}
\usepackage{subcaption}
\usepackage{footnote}
\usepackage{url}
\usepackage{tcolorbox}
\usepackage[para]{footmisc}
\usepackage{array,ragged2e}
\usepackage{multicol}
\usepackage{listings}
\usepackage{balance}
\usepackage{tipa}
\usepackage{rotating}

\definecolor{Gray}{gray}{0.9}

\def\BibTeX{{\rm B\kern-.05em{\sc i\kern-.025em b}\kern-.08em
    T\kern-.1667em\lower.7ex\hbox{E}\kern-.125emX}}
\begin{document}

\newcommand{\ie}{\emph{i.e.,}\xspace}
\newcommand{\eg}{\emph{e.g.,}\xspace}
\newcommand{\etc}{etc.\xspace}
\newcommand{\etal}{\emph{et~al.}\xspace}
\newcommand{\secref}[1]{Section~\ref{#1}\xspace}
\newcommand{\figref}[1]{Fig.~\ref{#1}\xspace}
\newcommand{\listref}[1]{Listing~\ref{#1}\xspace}
\newcommand{\tabref}[1]{Table~\ref{#1}\xspace}
\newcommand{\tool}[1]{{\sc #1}\xspace}
\newcommand{\commitref}[1]{$\mathtt{#1}$\xspace}
\newcommand{\RQ}[1]{RQ$_{#1}$\xspace}
\newcommand{\videogame}[1]{videogame$_{#1}$\xspace}

\newcommand{\apprach}{\textsc{...}\xspace}
\newcommand{\approachfullname}{{..}\xspace}

\newcommand\SIMONE[1]{\textcolor{red}{\nb{SIMONE}{#1}}}
\newcommand\ROCCO[1]{\textcolor{red}{\nb{ROCCO}{#1}}}
\newcommand\GABRIELE[1]{\textcolor{red}{\nb{GABRIELE}{#1}}}
\newcommand\EMANUELA[1]{\textcolor{red}{\nb{EMANUELA}{#1}}}
\newcommand\TODO[1]{\textcolor{red}{\nb{TODO}{#1}}}
\newcommand{\REV}[1]{#1\xspace}

\definecolor{gray50}{gray}{.5}
\definecolor{gray40}{gray}{.6}
\definecolor{gray30}{gray}{.7}
\definecolor{gray20}{gray}{.8}
\definecolor{gray10}{gray}{.9}
\definecolor{gray05}{gray}{.95}

\newlength\Linewidth
\def\findlength{\setlength\Linewidth\linewidth
  \addtolength\Linewidth{-4\fboxrule}
  \addtolength\Linewidth{-3\fboxsep}
}
\newenvironment{examplebox}{\par\begingroup
  \setlength{\fboxsep}{5pt}\findlength
  \setbox0=\vbox\bgroup\noindent
  \hsize=0.95\linewidth
  \begin{minipage}{0.95\linewidth}\normalsize}
  {\end{minipage}\egroup
  \textcolor{gray20}{\fboxsep1.5pt\fbox
    {\fboxsep5pt\colorbox{gray05}{\normalcolor\box0}}}
  \endgroup\par\noindent
  \normalcolor\ignorespacesafterend}
\let\Examplebox\examplebox
\let\endExamplebox\endexamplebox

\newenvironment{resultbox}{\vspace{0.2cm}\par\begingroup
  \setlength{\fboxsep}{5pt}\findlength
  \hspace{-0.5cm}
  \vspace{0.1cm}
  \setbox0=\vbox\bgroup\noindent
  \hsize=0.95\linewidth
  \begin{minipage}{0.95\linewidth}\normalsize}
  {\end{minipage}\egroup
  \textcolor{gray20}{\fboxsep1.5pt\fbox
    {\fboxsep5pt\colorbox{white}{\normalcolor\box0}}}
  \endgroup\par\noindent
  \normalcolor\ignorespacesafterend}
\let\Examplebox\examplebox
\let\endExamplebox\endexamplebox

\newboolean{showcomments}
\setboolean{showcomments}{true}

\ifthenelse{\boolean{showcomments}}
  {\newcommand{\nb}[2]{
    \fbox{\bfseries\sffamily\scriptsize#1}
    {\sf\small$\blacktriangleright$\textit{#2}$\blacktriangleleft$}
   }
  }
  {\newcommand{\nb}[2]{}
  }

\title[Towards Using Gameplay Videos for Detecting Issues in Video Games]{Towards Using Gameplay Videos for \\Detecting Issues in Video Games}
\titlenote{This study was accepted at the MSR 2022 Registered Reports Track.}

\author{Emanuela Guglielmi}
\affiliation{
    \institution{STAKE Lab\\University of Molise} \country{Italy}
}

\author{Simone Scalabrino}
\affiliation{
    \institution{STAKE Lab\\University of Molise} \country{Italy}
}

\author{Gabriele Bavota}
\affiliation{
    \institution{SEART @ Software Institute\\ Università della Svizzera italiana} \country{Switzerland}
}

\author{Rocco Oliveto}
\affiliation{
    \institution{STAKE Lab\\University of Molise} \country{Italy}
}
%

\newcommand{\approach}{GELID\xspace}
\newcommand{\approachlong}{GamEpLay Issue Detector\xspace}

\begin{abstract}
\textit{Context.} The game industry is increasingly growing in recent years. Every day, millions of people play video games, not only as a hobby, but also for professional competitions (\eg e-sports or speed-running) or for making business by entertaining others (\eg streamers). The latter daily produce a large amount of gameplay videos in which they also comment live what they experience. Since no software and, thus, no video game is perfect, streamers may encounter several problems (such as bugs, glitches, or performance issues). However, it is unlikely that they explicitly report such issues to developers. The identified problems may negatively impact the user's gaming experience and, in turn, can harm the reputation of the game and of the producer.
\textit{Objective.} We aim at proposing and empirically evaluating \approach, an approach for automatically extracting relevant information from gameplay videos by (i) identifying video segments in which streamers experienced anomalies; (ii) categorizing them based on their type and context in which appear (\eg bugs or glitches appearing in a specific level or scene of the game); and (iii) clustering segments that regard the same specific issue.
\textit{Method.} We will build on top of existing approaches able to identify videos that are relevant for a specific video game. These represent the input of \approach that processes them to achieve the defined objectives. We will experiment \approach on several gameplay videos to understand the extent to which each of its steps is effective.

\end{abstract}

\ccsdesc[500]{Software and its engineering~Software evolution}
\ccsdesc[500]{Software and its engineering~Maintaining software}
\ccsdesc[300]{Software and its engineering~Software defect analysis}

\keywords{video games, gameplay videos, mining software repositories}

\maketitle


\section{Introduction} \label{sec:intro}
Video games are becoming an increasingly important form of expression in Today's culture. Their sociological, economic, and technological impact is well recognized in the literature \cite{jones2008meaning} and their wide diffusion, particularly among the younger generations, has contributed to the growth of the gaming industry in several directions. Playing video games is progressively becoming a work for many: Some play for professional competitions (\eg in e-sports or speed-running), while others play to entertain others (\eg streamers) especially on dedicated platforms such as Twitch\footnote{\url{https://twitch.tv}}.
Besides all challenges that are common to software systems, developing and maintaining video games poses additional difficulties related to complex graphical user interfaces, performance requirements, and higher testing complexity. Concerning the latter point, games tend to have a large number of states that can be reached through different choices made by the player. In such a context, writing automated tests is far from trivial due to the need for an ``intelligent'' interaction triggering the states exploration. Even assuming such ability to explore the game space, determining what the correct behavior is in a specific state usually requires human assessment, with the exception of bugs causing the game to crash. Finally, additional complexity is brought by the non-determinism that occurs in games because of multi-threading, distributed computing, artificial intelligence and randomness injected to increase the difficulty of the game \cite{murphy2014cowboys}. 

Because of the few automated approaches available for quality control in video game development \cite{santos2018computer}, many games are released with unknown problems that are revealed only once customers start playing \cite{truelove2021we}. Since many streamers daily publish hours of gameplay videos, it is very likely that some of them experience such issues and leave traces of them in the uploaded videos. An example is available in \cite{youtubecrash}: The game crashes as soon as the player performs a specific action. 
The large amount of publicly available gameplay videos, therefore, might be a goldmine of information for developers. Indeed, such videos not only contain information about which kinds of issues affect a video game, but they also provide examples of interactions that led to the issue in the first place, allowing its reproduction. 
In their seminal work on this topic, Lin \etal \cite{lin2019identifying} defined an approach able to automatically identify videos containing bug reports. However, such an approach mostly relies on the video metadata (\eg its length) and it is not able to pinpoint the specific parts of the video in which the bug is reported. This makes it unsuitable as a reporting tool for game developers, especially when long videos, which are not uncommon, are spot as bug-reporting.

Our goal is to introduce \approach (\approachlong), an automated approach that aims at automatically extracting meaningful segments of gameplay videos in which streamers report issues and hierarchically organize them. 
Given some gameplay videos as input, \approach (i) partitions them in meaningful segments, (ii) automatically distinguishes informative segments from non-informative ones by also determining the type of reported issue (\eg bug, performance-related), (iii) groups them based on the ``context'' in which they appear (\ie whether the issue manifests itself in a specific game area), and (iv) clusters fragments related to the same specific issue (\eg the game crashes when a specific item is collected). 
In this registered report, \REV{we present the plan of an exploratory study to empirically evaluate \approach}. We first aim to extract training data for the machine learning model we plan to use to categorize segments. To this end, we will use the approach by Lin \etal \cite{lin2019identifying} to identify candidate videos from which we can manually label segments in which the streamer is reporting a bug. Then, we will run \approach on a set of real gameplay videos and validate its components by manually determining to what extent: (i) the extracted segments are usable, by annotating their \textit{interpretability} (\ie they can be used as standalone videos) and \textit{atomicity} (\ie they can not be further split); (ii) the category determined by \approach is correct, by computing typical metrics used to evaluate ML models (\eg accuracy and AUC); (iii) the clusters identified in terms of context and specific issues are valid, by performing both a quantitative (\eg using metrics such as the MoJoFM \cite{wen2004effectiveness}) and qualitative analysis of the obtained clusters. Finally, we will validate \approach as a whole and, specifically, the usefulness of the information it provides by running a survey with developers.

\section{Background and Related Work} \label{sec:Related}

\begin{table}
  \centering
  \caption{Mapping between types of issues identified by \approach and categories from the taxonomy by Truelove \etal \cite{truelove2021we}.}
  \label{tab:taxonomy}%
  \resizebox{0.9\linewidth}{!}{
    \begin{tabular}{l p{3cm} l}
    \toprule
    \textbf{Issue Type} & \textbf{Description} & \textbf{Categories \cite{truelove2021we}}\\
    \midrule
    \multirow{11}{*}{\textbf{Logic}} & 
    \multirow[t]{11}{*}{
    \parbox[t]{3cm}{Issues related to the game logic, regardless of how information is presented to the player.
    }} & 
        Object Persistence \\
    & & Collision of Objects\\
    & & Inter. btw. Obj. Prop.\\
    & & Position of Object\\
    & & Context State\\
    & & Crash\\
    & & Event Occurrence\\
    & & Interrupted Event\\
    & & Triggered Event\\
    & & Action\\
    & & Value\\
    \midrule
    
    \multirow{6}{*}{\textbf{Presentation}} & 
    \multirow[t]{6}{*}{
    \parbox[t]{3cm}{Issues related to the game interface (graphical- or audio- related).
    }} & 
        Game Graphics\\
    & & Information\\
    & & Bounds\\
    & & Camera\\
    & & Audio\\
    & & User Interface\\ 
    \midrule
    
    \multirow{2}{*}{\textbf{Balance}} & 
    \multirow[t]{2}{*}{
    \parbox[t]{3cm}{
    Detrimental aspects in terms of ``fun''.
    }} & 
        Artificial Intelligence\\
    & & Exploit\\
    \midrule
    
    \multirow{2}{*}{\textbf{Performance}} & Performance-related issues (\eg FPS drops). & Implem. Response\\
    \bottomrule
    \end{tabular}%
    }
\end{table}%

Several works have focused the attention on the quality assurance of video games analyzing the differences between traditional software development and video games development \cite{murphy2014cowboys, santos2018computer}. Given the goal of our approach, we mainly focus our discussion on approaches aimed at mining and manipulating gameplay videos for different goals. Also, since \approach aims at automatically categorizing video segments, we also discuss existing taxonomies of video game issues we will use as a starting point to define our categories.

\subsection{Mining of Gameplay Videos}
Some works targeted the automated generation of a comprehensive description of what happens in gameplay videos (\ie game commentary). Examples of these works are the framework by Guzdial \etal \cite{guzdial2018towards} and the approach presented by Li \etal \cite{li2019end} modeling the generation of commentaries as a sequence-to-sequence problem, converting video clips to commentary. On the same line of research, Shah \etal \cite{shah2019automated} presented an approach to generate automatic comments for videos by using deep convolutional neural networks.

The main goal of our approach, however, is to detect issues in gameplay videos. To the best of our knowledge, the only work aimed at achieving a similar goal is the one by Lin \etal \cite{lin2019identifying}. The authors investigate whether videos in which the streamer (player) experiences faults can be automatically identified from their metadata. They observe that na\"{\i}ve approaches based on keywords matching are inaccurate. Therefore, they propose an approach that uses a Random Forest classifier \cite{ho1995random} to categorize gameplay videos based on their probability of reporting a bug.
Lin et al. \cite{lin2019identifying} rely on Steam\footnote{\url{https://steamcommunity.com/}} to find videos related to specific games. While Steam is mainly a marketplace for video games, it also allows users to interact with each other and share videos. On a daily basis, for 21.4\% of the games on Steam, users share 50 game videos, and a median of 13 hours of video runtime \cite{lin2019identifying}. 
Hence, manually watching long gameplay videos classified as buggy still requires a considerable manual effort.
\approach aims at reducing the effort required by developers by segmenting videos and augmenting the provided information, by including also (i) the type of issue found, (ii) the context (\ie area of the game) in which it occurred, and (iii) other segments in which the same issue was reported (possibly from different videos).

\subsection{Taxonomies of Video Game Issues}
Video games can suffer from a vast variety of problems. Lin \etal \cite{lin2019identifying} do not distinguish among the types of issues reported in the videos identified as ``bug reporting'', while this is one of our goals. 

A recent taxonomy of issues in video games by Truelove \etal \cite{truelove2021we} (which extends the one by Lewis \etal \cite{lewis2010went}) reports 20 different kinds of issues. We will use such a taxonomy as a base to define the labels we want to assign to the video segments. However, using all such labels might be counterproductive since it is likely to observe a long-tail distribution (i.e., a few types of issues appear in most of the video fragments, while several other issues are quite rare or do not even appear). Therefore, starting from such a taxonomy, we define macro-categories by clustering similar fine-grained categories. We identified four labels, as reported in \tabref{tab:taxonomy}: \textit{Logic}, \textit{Presentation}, \textit{Balance}, and \textit{Performance}. 

 
\section{\approach in a nutshell} \label{sec:approach}

\approach takes as input a set of gameplay videos related to a specific video game, and it returns a hierarchy of segments of gameplay videos organized on three levels: (i) context (\eg level or game area), (ii) issue type (\eg bug or glitch), and (iii) specific issue (\eg game crashes when talking to a specific non-player character).

\begin{figure} 
	\centering\includegraphics[width=\linewidth]{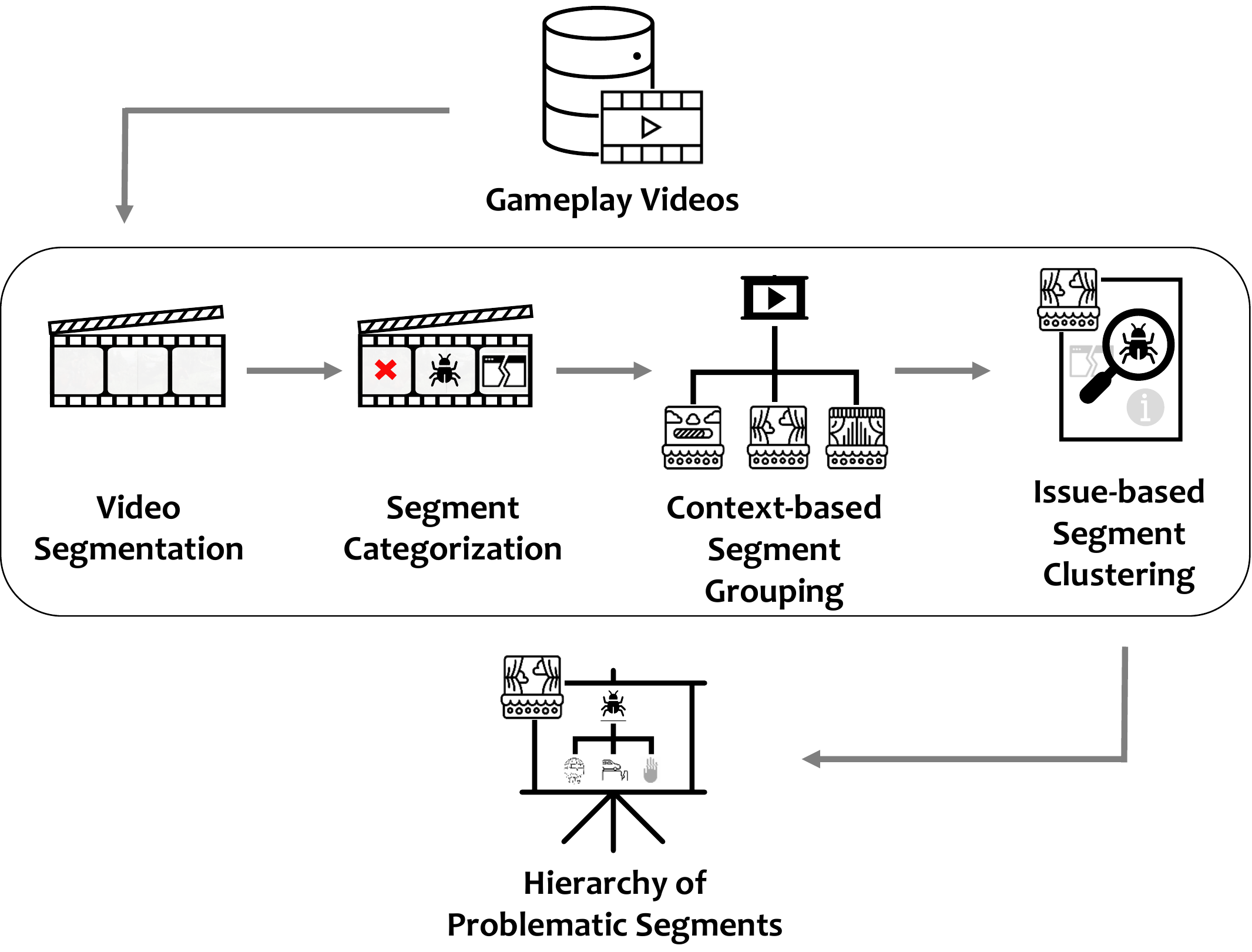}
	\captionsetup{justification=centering}
	\caption{The workflow of \approach}
	\label{fig:workflow}
\end{figure}

\figref{fig:workflow} shows an overview of the \approach workflow. We describe below in more detail the main steps of \approach.

\subsection{Video Segmentation}
The first step of \approach consists in partitioning the video in meaningful segments that can be later analyzed as standalone shorter videos. In other words, \approach aims at finding a set of ``cut points'' in the video.
In the computer vision literature, a similar problem is referred to as ``shot transitions detection''. The aim is to detect sudden changes in the video content. An example of approaches defined to solve such a problem is the one introduced by Tang \etal \cite{tang2018fast}.
Video-related information, however, might not be sufficient to find cuts in gameplay contents. For example, if the game crashes and a shot transition detection approach is used to cut the video, the second in which the crash happens would probably be selected for segmentation. The streamer, however, might need a few seconds to react to such an event by commenting what happened providing useful information for the game developers. Thus, by using shot transitions as cut points, the spoken content related to the issue might be erroneously put in the subsequent segment. To solve this problem, \approach relies on a blended video- and subtitle-based cut point detection algorithm. First, we detect shot transitions using the technique defined by Tang \etal \cite{tang2018fast}. Then, we shift each shot transition detected by $k$ seconds (where $k$ is a parameter that we will tune as part of our experiments), to account for the reaction time of the streamer. Finally, for each shifted shot transition, we will set a cut point at the moment in which the sentence pronounced by the streamer terminates, based on subtitles data.
For example, let us consider the case in which a shot transition is detected at the minute 13:45 (mm:ss) with $k = 5$. \approach first shifts the detected shot transition at 13:50 (13:45 + $k$). Assuming that at 13:50 the streamer is in the middle of a sentence, and they finish pronouncing it at 14:05, \approach sets a cut point at 14:05.

\subsection{Segment Categorization} 
In this second step, \approach aims at categorizing segments based on their content. \approach considers five labels: One for \textit{non-informative} segments (\ie the ones not reporting issues), and four for \textit{informative} segments (\ie the ones reported in \tabref{tab:taxonomy}). Non-informative segments are discarded and not considered in the next steps.

Previous work successfully used machine-learning to solve similar classification problems in the context of mobile app reviews \cite{chen2014ar, scalabrino2017listening}. Such approaches mainly rely on textual features. In our context, we also have video and audio, which could help to correctly classify the segments. For example, segments with no video might be more likely to be \textit{non-informative}, even if a comment by the player is present. Therefore, we include in \approach also video- and audio-based features. We will test different machine-learning algorithms to check which one allows to obtain the best results. Specifically, we will evaluate Random Forest \cite{ho1995random}, Logistic Regression \cite{ozkale2018logistic}, and Neural Network \cite{franklin2005elements}.
\REV{It is worth noting that, regardless of how precisely it is detected, a segment may show several issues at a time. For example, the game may start lagging and, at the same time, graphical glitches appear. At this stage, we do not handle segments reporting more than a issue at a time. In other words, we assume that each segment has exactly one label.}

\subsection{Context-based Segment Grouping} 
After having collected and categorized segments that contain anomalies, we group them accordingly to their \textit{context}.
With ``context'' we refer to the part of the game (\eg a specific game level or area) in which the anomaly occurred. This may be helpful to provide the videos to the team in charge of the development of that specific part of the game.

Such a step is important for two reasons: (i) Developers analyzing hundreds of videos related to a specific game may experience information overload and this, in turn, would reduce the effectiveness of the video segments filtering step; (ii) Knowing the context in which more anomalies occur allows the developer to identify where attention needs to be focused to improve the gaming experience.
To achieve this goal, we will rely on video information: The assumption is that videos with similar frames regard, most likely, the same context. Since the number of scenes is not necessarily known \emph{a priori}, we will use a non-parametric clustering technique. Specifically, we will experiment with DBSCAN \cite{ester1996density}, OPTICS \cite{ankerst1999optics}, and Mean Shift \cite{fukunaga1975estimation}.
\REV{We use as the distance metric for clustering a measure based on the image similarity \cite{ojeda2012measure, shechtman2007matching}. Specifically, given two videos, A and B, we first detect the key-frames in both the videos; Then, we compute the image similarity of each key-frame of A with each key-frame of B. Finally, we compute the average.}

\subsection{Issue-based Segment Clustering}
A set of video segments of the same kind (\eg bugs) and reported in the same context might still be hard to manually analyze for developers. For example, if 100 segments report bugs for a given level, developers need to manually analyze all of them. It might be the case, however, that most of them report the same specific bug (\eg a game object disappears). To reduce the effort required to analyze such information, we cluster segments reporting the same specific issue. This would allow developers to analyze a single segment for each cluster to have an overview of the problems affecting the specific area of the game. To achieve this goal, we will rely on both textual and image-based features, and we will use non-parametric clustering to create homogeneous groups. Textual features can help grasping the broad context (\eg objects disappearing or anomalous dialogues). Image-based features can help finding visually similar problems (\eg in the case of glitches). Similarly to the previous step, we will test several non-parametric clustering algorithms, such as DBSCAN \cite{ester1996density}, OPTICS \cite{ankerst1999optics}, and Mean Shift \cite{fukunaga1975estimation}.

\section{Research Questions} \label{sec:research}
The \textit{goal} of our study is to understand to what extent \approach allows to extract meaningful information from gameplay videos.

To achieve this goal, our study is steered by the following research questions (RQs).\begin{resultbox}
 \textbf{\RQ{1}}: \textit{How meaningful are the gameplay video segments extracted by \approach?}
\end{resultbox}
The first RQ aims at evaluating the quality of the segments extracted by \approach from gameplay videos in terms of their \textit{interpretability} and \textit{atomicity}.

\begin{resultbox}
 \textbf{\RQ{2}}: \textit{To what extent is \approach able to categorize gameplay video segments?}
\end{resultbox}
With this second RQ we want to understand which features and which classification algorithm allow to train the best model for categorizing gameplay video segments (second step of \approach), and what is the accuracy of such a model.

\begin{resultbox}
 \textbf{\RQ{3}}: \textit{What is the effectiveness of \approach in grouping gameplay video segments by context?}
\end{resultbox}
In the third RQ we aim to define the best clustering algorithm for grouping segments based on the game context (third step of \approach), and how effective is such an algotithm in absolute terms.

\begin{resultbox}
 \textbf{\RQ{4}}: \textit{What is the effectiveness of \approach in clustering gameplay video segments based on the specific issue?}
\end{resultbox}
Similarly to \RQ{3}, with this last RQ, we want to understand which features and clustering algorithm allow to achieve the best results for clustering segments based on the specific issue (fourth step of \approach), and how effective is such an algorithm in absolute terms.

\begin{resultbox}
 \textbf{\RQ{5}}: \textit{To what extent is the information provided by \approach useful to practitioners?}
\end{resultbox}
With this last RQ, we want to evaluate \approach as a whole. Specifically, we want to understand the perceived usefulness of the information provided by \approach, and which pieces of information are more relevant than others according to practitioners. 

\section{Datasets}
\label{sec:datasets}
To answer our RQs and validate the defined approach, we will rely on gameplay videos from YouTube. While other platforms, even more video game-oriented, could be used (\eg Twitch), YouTube provides APIs for searching videos of interest and it also allows to download videos including subtitles, which are required by \approach. While subtitles can be automatically generated when the video lacks them, the results could be noisy and, in this phase, we aim at evaluating \approach assuming high-quality input data.

In our study, we will collect three datasets. The first one, composed by video segments, will be used for training the supervised model used in step 2 of \approach (\ie segment categorization). The second one, composed by complete videos, will be used for evaluating the single components of \approach and answer \RQ{1-4}. The third one, composed by the output of \approach on a set of gameplay videos of a specific video game, will be used for evaluating \approach as a whole with practitioners. \REV{We will publicly release all the datasets to foster future research in this field}.

\subsection{Training Data}
We will use the APIs provided by YouTube to select a random sample of gameplay videos \REV{in English} from a diverse set of video games. While our premise is that several gameplay videos report issues, we also expect that the issues-reporting videos represent a minority of the entire gameplay videos population (thus the relevance of our approach). Therefore, to support the construction of the dataset containing training data for the categorization step, we will use the approach defined by Lin \etal \cite{lin2019identifying} and consider only videos identified as issue-reporting. Since our approach works at segment-level, we expect to collect a sufficient number of non-informative segments from videos reporting issues. Thus, the issue-reporting videos will provide training data for all categories of segments (\ie non-informative plus the four informative sub-categories). Also, the approach we will use is not perfect, so we will end up analyzing videos not reporting bugs anyway.
\REV{At this stage, we will only include videos with subtitles since \approach relies on NLP-based features computed on them. Some YouTube videos have manually-defined subtitles, while others have automatically generated ones. We include both of them. Indeed, while it is possible that the second category contain errors, this risk also exists in manually generated ones. Also, the general quality of the subtitles generated by YouTube is generally quite high for the English language.}

A human evaluator will manually split each video into meaningful segments, and at least two human evaluators will label each segment as \textbf{logic}, \textbf{presentation}, \textbf{balance}, \textbf{performance}, or \textbf{non-informative}. At this stage, our goal is to collect at least 1,000 labeled segments. \REV{To achieve this goal, we will run the previously mentioned process on batches of 500 videos at a time, until we collect the desired target number of segments. For each batch, we will randomly sample YouTube videos matching a generic query (\eg ``gameplay'') to ensure that the final sample is representative enough. We choose to include at least 1,000 segments in the training set because (i) such a number would probably allow us to appropriately train the model, and (ii) it is sufficient to have a representative set of segments. Indeed, assuming an infinite population (\ie we have an indefinitely high number of segments) and a 95\% confidence level, a sample of 1,000 segments allows us to have a 3.1\% margin of error, which we find acceptable in this context.}
\REV{To understand whether it is feasible to collect such an number of segments, we run a preliminary analysis. We searched for YouTube videos regarding the popular game Grand Theft Auto 5 (GTA 5). We found a total of 9,460,000 gameplay videos results. Given such a high number of results for a single video game, we believe that finding 1,000 segments is feasible.}

\REV{Finally, it is possible that the training set contains a few instances for some categories of issue types. For this purpose, if the training set is unbalanced, we will use oversampling techniques (\eg SMOTE \cite{SMOTE}) to generate synthetic instances for underrepresented categories.}

\subsection{Components Validation Data}
To select videos on which we will validate the single components of \approach, we will first need to select a small set of specific video games. Indeed, the third and fourth steps of \approach are reasonable only when segments from the same video game are given. To select the video games to use, we will rely on the information available on Steam, one of the largest video game marketplaces \cite{toy2018large}. We will select three video games that are both popular (\eg for which many gameplay videos exist) and that had several reported issues (\eg for which \approach gives the best advantage). More specifically, we will select video games with many downloads, low review scores and many patches. Then, we will collect a random sample of gameplay videos related to each selected video game from YouTube. \REV{Also in this case, we will select only English videos with subtitles (either manually added or automatically generated).}
\REV{Since we will test different machine-learning techniques (both for categorization and clustering), we will need to tune their respective hyper-parameters. To this aim, we will use 10\% of the data acquired at this stage as \textit{evaluation set}, and the remaining 90\% as test set.} 

\subsection{Approach Validation Data}
To select videos on which we will validate \approach as a whole, we will first select a video game on Steam and extract a set of gameplay videos from YouTube about such a game. To do this, we will use the exact same approach used to build the previously described dataset, but with a fourth game. Then, we will feed \approach with such videos. \approach, in turn, will provide information about (i) contexts (\ie area of the game), (ii) issue types (\eg logic or presentation issue), and (iii) specific issue. At this stage, we will use the best configuration of \approach, based on the findings of \RQ{1-4} (\eg the most accurate categorization model).
We will ask practitioners to evaluate the information provided by \approach, as detailed in \secref{sec:execution}.

\begin{table}
  \centering
  \caption{Questions for the survey we will run to answer \RQ{5}.}
  \label{tab:survey}%
  \resizebox{\linewidth}{!}{
    \begin{tabular}{l p{8cm} l}
    \toprule
    \multirow{10}{*}{\begin{rotate}{90} Pre-quest. \end{rotate}}
     & \textbf{Question} & \textbf{Type of response} \\
    \midrule
    & Full name & Text \\
    & Email address & Text\\
    & Education & Multiple selection (\eg graduate)\\
    & Role & Multiple selection (\eg tester)\\
    & Years of experience & Number \\
    \midrule
    \multirow{28}{*}{\begin{rotate}{90} Questionnaire \end{rotate}}
    & \multicolumn{2}{c}{\textbf{Context information}}\\
    & To what extent is context summary information useful? & 5-point Likert scale \\
    & To what extent is the ability to navigate segments from contexts useful? & 5-point Likert scale\\
    & Please justify your answers & Open response\\
    \cmidrule{2-3}
    
    & \multicolumn{2}{c}{\textbf{Issue category information}}\\
    & To what extent is issue category summary information useful? & 5-point Likert scale \\
    & To what extent is the ability to navigate segments from issue categories useful? & 5-point Likert scale \\
    & Please justify your answers & Open response\\
    \cmidrule{2-3}
    
    & \multicolumn{2}{c}{\textbf{Specific issue information}}\\
    & To what extent is specific issue summary information useful? & 5-point Likert scale \\
    & To what extent is the ability to navigate segments from clusters of specific issues useful? & 5-point Likert scale \\
    & Please justify your answers & Open response\\
    \cmidrule{2-3}
    
    & \multicolumn{2}{c}{\textbf{Segments information}}\\
    & How useful would the segments be to understand issues? & 5-point Likert scale\\
    & How useful would the segments be to reproduce issues? & 5-point Likert scale \\
    \cmidrule{2-3}
    
    & \multicolumn{2}{c}{\textbf{Global evaluation}}\\
    & How useful would \approach be during the closed testing phase? & 5-point Likert scale \\
    & How useful would \approach be  during the beta testing phase? & 5-point Likert scale \\
    & How useful would \approach be  during the production phase? & 5-point Likert scale \\
    & What are the strengths of \approach? & Open response \\ 
    & What are the weaknesses of \approach? & Open response \\ 
    & Additional comments (optional)  & Open response\\
    \bottomrule
    \end{tabular}%
    }
\end{table}%

\section{Execution Plan}
\label{sec:execution}
\begin{figure*}
 \centering\includegraphics[width=\linewidth]{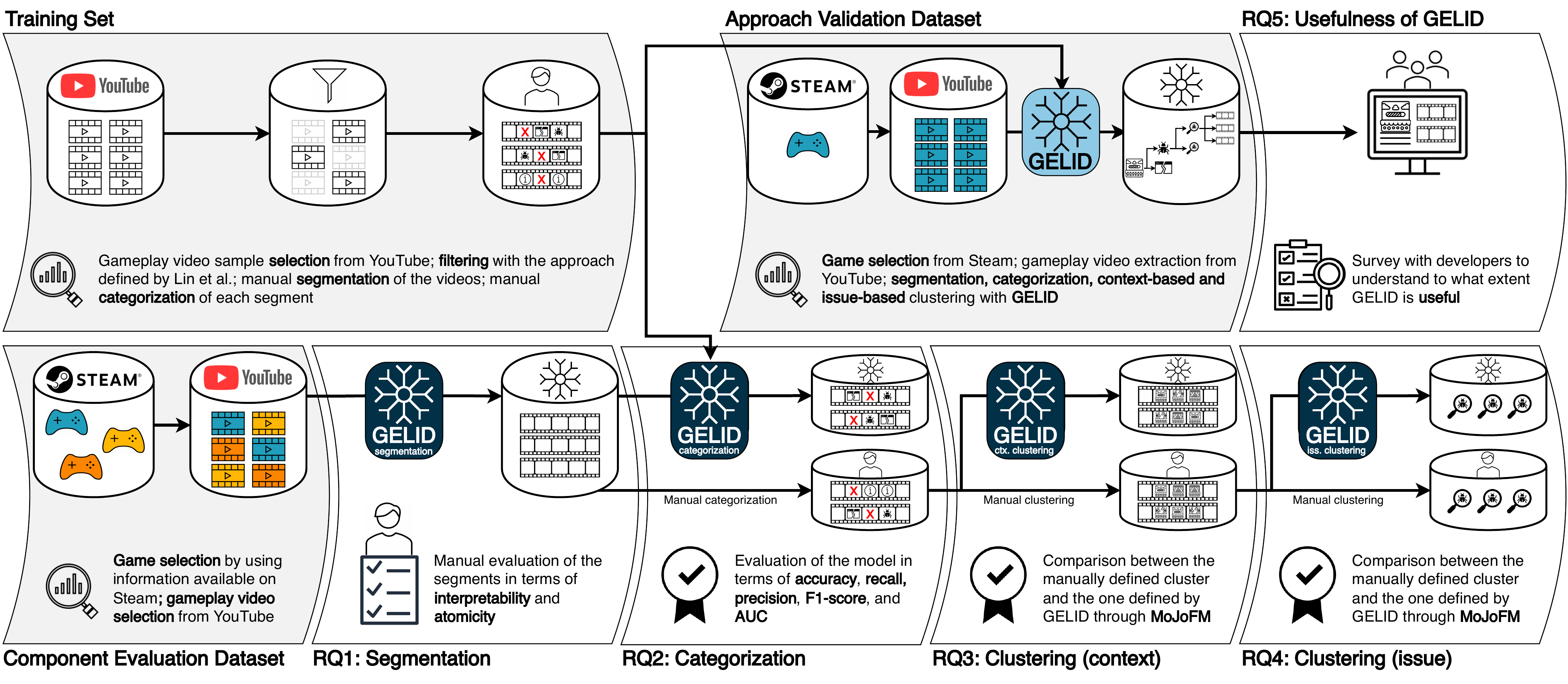}
  \captionsetup{justification=centering}
  \caption{Summary of the study design.}
  \label{fig:plan}
\end{figure*}

We will run the following plan to answer our research questions and conduct the study. We summarize the execution plan in \figref{fig:plan}.

\subsection{Research Method for \RQ{1}}
To answer \RQ{1}, we will evaluate the technique we defined with different values of $k$ (streamer reaction times). Specifically, we will instantiate our approach with $k$ in the set $\{0, 5, 10\}$ seconds. 

We will evaluate the segments detected by each variant of our approach in terms of their (i) \textit{interpretability} (\ie it is possible to watch the segment and acquire all the information needed to understand what has been experienced by the streamer) (ii) the \textit{atomicity} (\ie it is not possible to further split the segments). Such aspects are complementary: It would be possible to maximize the \textit{interpretability} by creating few segments (\eg just one for the whole video); this, however, would result in lower \textit{atomicity} since the segments could be further divided into parts. Two human annotators will watch segments generated by each technique and manually annotate each segment in terms of its \textit{interpretability} and \textit{atomicity} on a 5-point Likert scale.  \REV{As for the first metric, we will ask the annotator to evaluate to what extent he/she can fully understand what is happening based only on the segment itself. As for atomicity, instead, we will ask annotators to assess if the segment can be further divided in additional standalone (fully interpretable) segments. The final score should be computed as 5 minus the number of additional standalone segments that can be further extracted, or 1 if more than four standalone segments are found.}
We will report the inter-rater reliability between the annotators by using the Cohen's kappa coefficient \cite{cohen1960coefficient}. Then, for each segment, we will compute the mean and median \textit{interpretability} and \textit{atomicity}.
Finally, we will compare the tested techniques in terms of such metrics using a Mann-Whitney U test \cite{mann1947test}, and adjusting the  $p$-values resulting for multiple comparisons using the Benjamini and Hochberg procedure \citep{benjamini1995controlling}. 
We will also report the effect size, using the Cliff's delta \citep{cliff1993dominance}, to understand the magnitude of differences observed. 
\REV{We run power analysis to define an adequate sample size that allows us to detect possible differences in terms of interpretability and atomicity among the three groups of segments (\ie with $k \in \{0, 5, 10\}$). First, we are interested in observing variations of at least 0.5 in the means of the groups for both the metrics; smaller variations would likely be practically irrelevant. To determine the expected standard deviation, we run 1,000 simulations in which we randomly assigned scores between 1 and 5, and we computed, for each simulated assignment, the standard deviation. We obtained values in the range $[1.28, 1.54]$. If we analyze 200 segments for each group, assuming the previously reported range of expected standard deviation values, we expect to achieve a power in the range $[90\%, 97\%]$. If we simulate the worst-case scenario in terms of standard deviation (\ie we have half observations with the lowest score and the other half with the highest score), we obtain a standard deviation of $\sim2$, which leads to a 71\% power. Even in such an unlikely scenario, we would have acceptable power. Therefore, we plan to use groups of 200 segments.}

\subsection{Research Method for \RQ{2}}
To answer \RQ{2}, we will train and test several approaches for categorization, using different sets of features and different machine learning algorithms to understand (i) which features are more relevant, and (ii) which machine learning technique allows to achieve the best results. For each tested algorithm (\ie Random Forest \cite{ho1995random}, Logistic Regression \cite{ozkale2018logistic}, Neural Network \cite{franklin2005elements}), we will define three models including (i) textual features only, extracted from the subtitles (\ie what the streamer says), (ii) video features only, extracted from the video (\ie what happens in the game), and (iii) all features together. In total, we will compare nine models. \REV{We will explore different types of NLP features, including bag-of-words \cite{zhang2010understanding} and word2vec \cite{rong2014word2vec}}.

To train each model, we will use the training set described in \secref{sec:datasets}. To define a \textit{test set}, instead, we will consider the segments obtained through the best segmentation technique, based on the results of \RQ{1}. Two human annotators will independently analyze each segment and manually label them as \textbf{logic}, \textbf{presentation}, \textbf{balance}, \textbf{performance}, or \textbf{non-informative}. The human annotators will discuss the cases of disagreement aiming at reaching consensus. If no consensus can be found, the segment will be discarded as ambiguous also for human evaluators.

For each model, we will report the global accuracy (\ie percentage of correctly classified instances) and, for each class, the achieved precision ($\frac{\mathit{TP}}{\mathit{TP} + \mathit{FP}}$), recall ($\frac{\mathit{TP}}{\mathit{TP} + \mathit{FN}}$), and AUC (Area Under the ROC Curve).

\subsection{Research Method for \RQ{3} and \RQ{4}}
To answer both \RQ{3} and \RQ{4}, we will test the following non-parametric clustering techniques: DBSCAN \cite{ester1996density}, OPTICS \cite{ankerst1999optics}, and Mean Shift \cite{fukunaga1975estimation}.
We will start from the test set defined to answer \RQ{2}. For each video game, two human annotators will group the segments based on the game context in which they appear. The human annotators will discuss conflicts aiming at reaching consensus. Segments on which consensus can not be reached will be discarded, similarly to \RQ{2}. We will use the ground-truth partition produced after this step to evaluate the previously-mentioned clustering techniques to answer \RQ{3} (more on this below).
Then, for each cluster and for each category of issues, we will further cluster segments according to the specific issue highlighted. We will use the same process used for building the ground-truth partition defined for answering \RQ{3}, and we will use such a partition to answer \RQ{4}.

For both the RQs, we will compare the models by using the MoJo eFfectiveness Measure (\textit{MoJoFM}) \cite{wen2004effectiveness}, a normalized variant of the MoJo distance. \textit{MoJoFM} is computed using the following formula:
$$
MoJoFM(A,B)  = 100 - (\frac{mno(A,B)}{max(mno(\forall E_{A}, B))} \times 100)
$$
\noindent where $mno(A,B)$ is the minimum number of \emph{Move} or \emph{Join} operations one needs to perform in order to transform a partition $A$ into a different partition $B$, and $max(mno(\forall \; E_{A}, B)$ is the maximum possible distance of any partition $A$ from any partition $B$. \textit{MoJoFM} returns 0 if partition $A$ is the farthest partition away from $B$; it returns 100 if $A$ is equal to $B$.

\subsection{Research Method for \RQ{5}}
To answer \RQ{5}, we will run a survey with at least five practitioners, aimed at collecting their opinions on the usefulness of the information provided by \approach. \REV{We preliminarly acquired the availability of five professional developers to achieve do this}. 
The survey will be composed of three steps. First, participants will complete a pre-questionnaire to acquire basic information. We report the specific questions we will ask in the top part of \tabref{tab:survey}. Second, we will ask them to freely browse for 15 minutes a web-app containing the information generated by \approach for a specific video game. To this aim, we will exploit the third dataset defined in \secref{sec:datasets}. Especially, participants will be able to: (i) view summary information about the contexts; (ii) browse contexts and view summary information about the categories of issues affecting them; (iii) browse issue categories and view summary information about clusters of video segments regarding specific issues; (iv) browse such clusters and watch video segments.
After this step, participants will answer specific questions about the perceived usefulness of \approach, as specified in the bottom part of \tabref{tab:survey}. We will report summary statistics about the responses and qualitatively analyze their comments to also provide insights about future research directions.

\section{Limitations, Challenges, and Mitigations} \label{sec:threats}

In this step we summarize the main limitations of our work and outline the mitigation strategies we use.

\textbf{Subjectivity of the Manual Analysis.} It is possible that the manually determined labels and partitions used to answer our RQs are not correct. To mitigate this limitation, all the manual evaluations are performed by two authors, and the results are discussed to reach consensus. 

\REV{\textbf{Evaluation Biases.} In the evaluation of \approach, we will explicitly select video games with many issues. This could result in a bias in the evaluation. It might be possible that we conclude that \approach works, while it works only on problematic games. However, we believe that \approach is most useful for such a category of video games}

\textbf{Incomplete Definition of Categorization Labels.} The definition of the labels used in the second step of \approach (\ie categorization) might be incomplete. It is possible that we do not consider some relevant categories of issues. To mitigate this limitation, we relied on a state-of-the-art taxonomy \cite{truelove2021we}.

\textbf{Ineffectiveness of the Features.} A key issue in implementing \approach consists in defining meaningful features for categorizing and clustering video segments. We will initially rely on features previously engineered in a similar context (\eg categorization and clustering of mobile app reviews \cite{chen2014ar, scalabrino2017listening}). A different set of features may lead to different, possibly better, results.

\REV{\textbf{Manual Analysis Effort.} We foresee a big amount of manual analysis to be performed to both (i) build a training set and (ii) answer our RQs. To address this challenge, we will likely involve other people in this process (\eg game players).}

\section{Conclusion} \label{sec:conclusion}
In recent years, there has been a growing interest in video games. During game development, many bugs go undetected prior to release because of the difficulty of fully testing all aspects of a video game.
We introduce \approach, a novel approach for detecting anomalies in video games from gameplay videos to support developers by providing them with useful information on how to improve their games. We presented the plan of our empirical study designed to understand to what extent \approach can provide meaningful information to developers.


\balance
\bibliographystyle{ACM-Reference-Format}
\bibliography{main}

\end{document}